\documentclass[conference]{IEEEtran}
\usepackage[dvips]{graphicx}

\ifCLASSINFOpdf
\else
\fi
\begin{document}
%
\title{An extendible User-Command Framework based on tagging system}

\author{\IEEEauthorblockN{Ajinkya G Kale}
\IEEEauthorblockA{IBM Software Group\\IBM India Pvt. Ltd.\\
Pune, India. 411006\\
Email: ajinkyakale@in.ibm.com}
\and
\IEEEauthorblockN{Ananth Chakravarthy}
\IEEEauthorblockA{IBM Software Group\\IBM India Pvt. Ltd.\\
Pune, India. 411006\\
Email: ananth\_chakravarthy@in.ibm.com}
\and
\IEEEauthorblockN{Nitin S Jadhav}
\IEEEauthorblockA{IBM Software Group\\IBM India Pvt. Ltd.\\
Pune, India. 411006\\
Email: nijadhav@in.ibm.com}}


%


\maketitle

\begin{abstract}
Memorizing the user commands has been a problem since long. In this study we try to propose solutions to overcome two problems -
the problem of selecting appropriate commands names during application development and the problem of memorizing these command names.
The proposed solution includes a framework in which the applications can plug into, to get their application commands and corresponding tags 
in to the new command execution application.We also propose a mechanism where user can generate her own set of tags for a command and share 
those with peers.
\end{abstract}

%
\IEEEpeerreviewmaketitle

\section{Introduction}
Natural Command names and the Initial learning of a command based system is a well studied area.
Landauer et al studied the question "would incorporating the novice's words into the language needed to operate an editing system make initial learning easier or the system more acceptable?" - in their research report [1]. They came up with a statistical survey which showed that the novices' claim that the system does not use their words is correct. The report claims that the probability that any two users would use the same verb in response to a particular text correction operation, is only 0.08. This shows that individual subjects were not even very self-consistent. They emphasize the fact that the likelihood that a given subject would use the same verb for both instances of an operation is 0.34.

Novice users do not use the same language or command names as system designer or developers do to describe operations. As a matter of fact one potential user does not even use the same set of words as another or even the same words on different occasions. This demonstrates a need to have an empirical analysis of the command names at design time.
But these empirical results may also vary depending upon factors like geography of users, their vocabulary, etc. This calls for a more intelligent solution to solve this "curse of memorization(of commands)". We try to solve this problem using a command-tagging mechanism.

So two basic questions raised during our study were : Is a language designed by users themselves more efficient than a language designed
by computer and human factors specialists[2] ? and Is it possible to provide an easy solution for novice users to use her own set of language for the commands and also be able to share the linguistics with other users.
Broadbent and Broadbent [4] show that the subjects were better at retrieving items from a data base with the descriptor terms they invented,
than with the terms that were supplied by the experimenter.However, that experiment did not address the specific issue of computer commands[2] 

The study presented here focuses on a tagging mechanism to help novice user with automated user-command execution systems.
The aim of the study is to present a tagging system for user-commands such that the natural language command grammar will be easy.
\\With this in mind the rest of the paper is organized as follows: Section 2 describes the problem setting. Section 3 presents a brief summary of the proposed solutions. Section 4 involves the implementation details of the proposed solutions. And we conclude with future scope of work in this area in section 5 and 6.

\section{Problem Setting}
The problem of learning the jargons of computing commands of an interactive system still remains acute. "Naming" has been of interests to philosophers, linguists and psychologists [6,7].
There is little semantic research on the psychological processes involved in the understanding and acquisition of the vocabularies of interactive computer systems[6,7,9].
In their study Barnard et al. show that ocassional users are faced with the task of understanding, learning and remembering new meanings for the words.Such considerations suggest that names which reflect users' own conceptions of command operations should facilitate learning[9].
The results of these studies suggest that alternative command names are likely to influence the novice user learning. These alternative command names will inturn be influenced by the study of the target users.

\subsection{Problem Statement}
The problem we focus on is the study of possible solutions for the "curse of memorization(of command names)". The problem deals with the understanding and acquisition of vocabularies pertaining to the interaction with computer systems. As every system designer cannot study the psychological aspect of the target users, there is a requirement of a generic solution for this problem. A solution which will allow the system to adapt unobtrusively from the usability point of view of the target audience. The rest of the paper deals with intuitive solutions which can be used as an aid to solve the preceding problem. The system has to easily let the user to personalize the command system according to her vocabulary and also allow her to share this personlized data with other users of her kind.
Our focus remains solely on the usability of a interactive computer system user who can adapt quickly with the linguistics involved in the computing operations.

\section{Solutions Proposed}
We will use tagging as a mean to overcome the user onus of command-names memorization to use any interactivie system which has a command vocabulary based operations.
Any new application will come along with its command-names and tags [section \_]. User can use any of the tags associated with the command to recall the command. Tags will be like a set of common words associated with the particular command name. They will be as good as alias for the command name. Using tagging, the words associated with a command are easy to recall and retrieve than the original user-command set. Tagging will be the backbone of the new proposed system which will help users to associate alias/tags of their choice to the commands.
There will be a many to one mapping from the tag-set to command-set.

The technicality of the implementation will have two aspects to it : \\
\\
1. The application developer will create a command  \\
\indent    -tags map in xml format which can be easily \\
\indent    consumed by the new user-command framework.  \\
\indent    This will be used to create a searchable db \\
\indent 		of the command-tags map.\\
2. The user will be able to intelligently and unobtrusively \\
\indent    create and store tags to the commands she issues.\\ 
\indent    This will be integrated with the command-tags db \\
\indent    and will be 'searchable' and 'sharable' in future.\\

\subsection{Consumable xml file format : command-tags map}
XML (Extensible Markup Language) is a general-purpose specification for creating custom markup languages.[a] It is classified as an extensible language, because it allows the user to define the mark-up elements. XML's purpose is to aid information systems in sharing structured data, to encode documents, and to serialize data [11].

During application development the application designer will provide commands and a basic set of tags to these commands. This mapping of tags and commands will be provided in XML format. The XML file will be the extension point of the new command framework. The command framework will parse this xml file and index the commands and their corresponding tags. These indexes will be maintained in a central repository which will be queried against during the user operations.
So whenever a user is using the command framework she will just have to type is a tag (or the actual command if she knows) to get the options of commands having that tag. These tags will be common language vocabulary so as there is no stress on the user brain to remember the application commands !

Tags will aid in the easy of the command use by not only providing easy means to recollect difficult commands but also by enabling a user to use her set of vocabulary (independent of the language) to remember her commands.

\subsection{User personalized tagging}
With the above proposed approach every user can personalize a command-based system. This personalization will be brought in using extensibility to the current already provided tag-command maps. User will be allowed to add her own tags to the existing command-tag maps. This means that the barricade of vocabulary and language dependency (for use of systems) will be eliminated.
With the extensibility provided, user will be able to harness this to share her tag-command maps with other users of similar tastes. This will help to ease the learning curve of the naive users who have similar background. Any advance user can easily share her maps with novices, on which the novice can develop their own set of tag-command layer for their aid.

Sharability will be a key aspect addressed here because it is one of the main issues while learning any new command system. User has to go through the manual for every basic operation initially or else she has to consult an advanced user every now and then to get familiarized with the command syntax.

\section{Implementation Details}

\begin{center}
\begin{figure}{ } \label{ex}
\resizebox{10cm}{10cm}{\includegraphics[0mm,0mm][600,500]{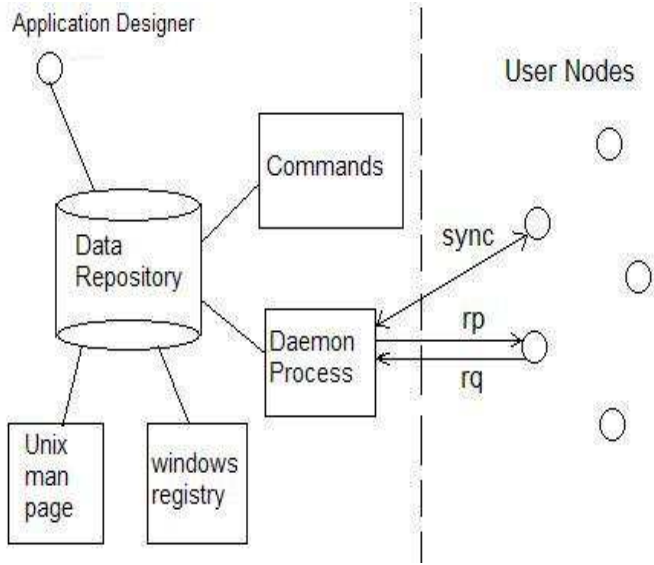} }
\caption{Architectural Block Diagram}
\end{figure}
\end{center}
\subsection{Architectural block diagram}

Figure 1, explains the generic block diagram of the proposed framework's architecture.\\
a) Outside world user will be accessing the command store through the daemon process. This will be in the form of request-response format.\\
b) Daemon process will also be looking over the user activities and storing them in the form of user-history.\\
c) Additionally the daemon process will be used to update the personalized user tags. These tags will be updated against the coressponding user command.\\
d) If the user wishes to synchronize her tags with any other user node in the user network then this synchronization can be bought about using the same daemon.\\
e) Application developer wil also play her part in creating the basic command-tags map ( which can be in xml format)\\
f) The data repository can be extended from the man pages store on unix based systems or from the windows registry on Windows.\\

\subsection{Example OS - Unix : Man Page Extension}
This section describes how the man tool available on the UNIX based systems could be used to extend and implement the tagging based command execution framework. This framework which is being described needs to meet the following criteria
1.	It should be able to provide a help context to execute a command. For example, if a user wants to remove a file, using the conventional shell on the UNIX system, he/she should be using the remove command. The framework should be able to prompt the user with the command name "rm" if he/she gives the related commands as "delete" or "remove" as the tags to be used for searching.
2.	It should be easily extendible without radical changes to the current infrastructure. 
3.	The knowledge captured in the various interactions with the user should be easily sharable across the machines. The knowledge that is captured can include the following aspects
o	The custom tags that can be added by the end user which can be shared with the other users 
o	The actual history of search tags that have been used by the end user and the actual command that has been used after trying various search options. 
o	The actual command that has been entered so that it can be captured as example usage. 

The following are the main modifications that are proposed to the current UNIX based systems. 
1.	The "man" binary is to be modified for various additional options. 
2.	The "man" page format is to be extended to support the new framework. 

The following section describes these two modifications in greater detail. 

\subsubsection{Changes to the man binary and man format extensions}
The conventional man binary on the UNIX platform works according to the following conditions. The environment variable MANPATH is used to look at the various locations that can be used to find a match to the command that is being used to find the help for. The man binary iterates over all the directories specified in the above path. The man binary understands the man files as they are written in a particular format.
The application developer who is writing the man files will be writing the man files that conforms to this format. As such the man binary just displays the information by interpreting this format. The primary rule for this format is that each module of information has to begin with a "keyword" from a possible list. Some of the commonly used keywords are .TH (Title Header), .SH (Section header), .P (a Paragraph) and so on. The proposed framework extends this format to add additional sections for this format. Some of the standard sections that can go in this format are NAME, SYNOPSIS, DESCRIPTION, OPTIONS, DIAGNOSTICS, and BUGS. The proposed framework adds the following additional sections "TAGS", "USAGE HISTORY" and "EXAMPLE USAGE". The man binary when modified as per the proposed framework would be able to identify these additional sections as well. As per the current man manual, any additional section can also be added to the man pages but they will simply displayed as information.  To summarize the following would be the major changes: 

TAGS - This man section will capture the tagging information of a command. As every command had a man page, this man document will have this TAGS section to enumerate all the tagging that has been gone into the system for the particular command. It will be a many to one mapping from tags to command. This section maybe then utilized by users to get command names using the tags.
eg : "man -tags delete" can return a list of command suggestions like "rm" , "rmdir" 

USAGE HISTORY - Usage History section will be a user specific section in which each user will have his own association of tags to the commands. This will be a personalization of the tagging system to aid the easy of issuing commands using user-specific tags (along with those provided as defaults). In this section there will be a pointer pointing to each users data store of personalized tag-command maps.

EXAMPLE USAGE - Example Usage will be captured by the daemon running on the system. It will capture all the cycle of a command execution process, starting from getting the commands from the tags (ie finding the right command) to executing the exact command. This will be useful in future in suggesting the user with the possible command usage example based on the history of the tags-commands combinations. Example usage will also be harnessed to suggest command combination likeliness ie if a particular command A was used in conjunction with command B in past then if user searches for tags on command A then he will be shown that in the past A was used in conjunction with B. 
\\
\subsubsection{Sharing of the Command-Tag maps}
Tags can be shared across using 2 techniques :

(a)A "synchronize" option can be added to the man parameters which will ask for the user node(over the network) with whom to synchronize. With this the other users tag history and usage history will be stored. More granular preferences can be provided which can help users to only publish specific tags while not publishing all of them.

(b)The daemon process can fetch and suggest the example usage of a command based on other users "usage history" and "example usage" sections.

\subsection{Goals for the changes in the man format}
In all the proposed changes to the "man" file format will adhere to the following goals :\\
1.	The ability of the modified man binary to detect the newly added sections\\
2.	The ability of the modified man binary to support new command line options to look into these various sections \\
3.	The ability of the man binary to look into the "TAGS" section to match a set of keywords as given by the user to peek into this section of every man entry. The set of keywords are passed using the new command line options identified in (2) above. \\
4.	The ability of the man binary to resolve a collection of matches based on the normalization of the keywords given by the user and the collection of man files that are having the given keywords as probable matches for them. \\
5.	The ability of the man binary to use a dictionary to do an extended lookup based on the keywords entered by the user provided there are no exact matches based on the TAGS section as proposed by the framework\\
6.	The ability of the man binary to suggest example usage basing on an additional command line option to be supported by the new framework. The example usage can be extracted from the new proposed section 'EXAMPLE USAGE". \\ 
7.	The ability of the man binary to correlate the extended usage of the command based on the keywords entered by the user in conjunction with other commands. For example the ability of the man binary to lookup the example usage of the current command along with another command like "grep" if additional command line options are passed by the user.\\
8.	The ability of the man binary to launch a daemon process to perform additional processing to capture user behavior and serve information sharing across machines or nodes in a network. \\
9.	The ability of the man daemon binary to scan the history file for a given user for sequences of the man command followed by the actual command that has been used by the user.\\
10.	The ability of the man daemon binary to capture usage immediately following a man command and the actual command line options used and correlate with the action command immediately issued by the user.\\ 
11.	The ability of the man binary to take inputs as TAGS from the user and store them as part of the "USAGE HISTORY" section. \\
12.	The man binary would not actually modify the man page itself but use a variable from the environment to use a pointer to a file that can be used to capture the semantics on a per user basis. This would allow the users from having their own space for capturing the tags associated with the man entry. \\
13.	The "USAGE HISTORY" section would only have the options as set by the man page creator to interpret this user specific file.\\
14.	The ability of the man daemon to pull a user specific set of tags that can be used to share the usage and tags based on the authentication data provided to the man daemon. \\

\section{Future Scope}

The whole framework can be extended to UI applications too such that power users can be provided with a command framework for the comman UI operations. (eg: Menu operations of a UI based application)
Each UI operation will be encapsulated into an operation object which will be mapped to a command. These commands will be used by the framework to perform the same UI operation using the command line interpreter.
In turn these UI operation commands can be tagged same as described above for command based systems.

\section{Conclusion}

The problems faced by end users as a part of their application learning (for command based interactive systems), gives rise to a need for a solution for the "command-memorizing" problem. The proposed solution can be a step towards addressing few of the naive users problems related to command-based systems.






\begin{thebibliography}{1}
\bibitem{12}
Scott A. Golder, Bernardo A. Huberman The Structure of Collaborative Tagging Systems. 
\\

\thanks{UNIX is a trademark of The Open Group in the United States, other countries, or both.
IBM is a trademark of IBM Corporation in the United States, other countries, or both.
Other company, product, or service names may be trademarks or service marks of others.}
\end{thebibliography}
%

\end{document}